\documentclass[%
 reprint,
 superscriptaddress,
 showpacs,
 amsmath,amssymb,
aip, jap,
]{revtex4-1}

\usepackage[dvipdfmx]{graphicx}
\usepackage{dcolumn}
\usepackage{bm}
\usepackage{subcaption}
\usepackage{tabularx,multirow,array,threeparttable}
\usepackage{color}
\usepackage[normalem]{ulem}

\newcommand{\eref}[1]{Eq.\,(\ref{#1})}
\newcommand{\figref}[1]{Fig.\,\ref{#1}}

\begin{document}


\title{
Elucidation of the atomic-scale mechanism of the anisotropic oxidation rate of 4H-SiC between the ($0001$) Si-face and ($000\overline{1}$) C-face by using a new Si-O-C interatomic potential
}

\author{So Takamoto}
\altaffiliation{Corresponding author}
\email{takamoto.so@fml.t.u-tokyo.ac.jp}
\affiliation{Department of Mechanical Engineering, School of Engineering, The University of Tokyo, 7-3-1 Hongo, Bunkyo-ku, Tokyo 113-8656, Japan}
\author{Takahiro Yamasaki}
\affiliation{International Center for Materials Nanoarchitectonics, National Institute for Materials Science, 1-1 Namiki, Tsukuba, Ibaraki 305-0044, Japan}
\author{Takahisa Ohno}
\affiliation{International Center for Materials Nanoarchitectonics, National Institute for Materials Science, 1-1 Namiki, Tsukuba, Ibaraki 305-0044, Japan}
\author{Chioko Kaneta}
\affiliation{Fujitsu Laboratories Ltd., 10-1 Morinosato Wakamiya, Atsugi, Kanagawa 243-0197, Japan}
\author{Asuka Hatano}
\affiliation{Department of Mechanical Engineering, School of Engineering, The University of Tokyo, 7-3-1 Hongo, Bunkyo-ku, Tokyo 113-8656, Japan}
\author{Satoshi Izumi}
\affiliation{Department of Mechanical Engineering, School of Engineering, The University of Tokyo, 7-3-1 Hongo, Bunkyo-ku, Tokyo 113-8656, Japan}

\date{\today}

\begin{abstract}
Silicon carbide (SiC) is an attractive semiconductor material for applications in power electronic devices.
However, fabrication of a high-quality SiC/SiO${}_2$ interface has been a challenge.
It is well-known that there is a great difference in oxidation rate between the Si-face and C-face, and that the quality of oxide on the Si-face is greater than that on the C-face.
However, the atomistic mechanism of the thermal oxidation of SiC remains to be solved.
In this paper, a new Si-C-O interatomic potential was developed to reproduce the kinetics of the thermal oxidation of SiC.
Using this newly developed potential, large-scale SiC oxidation simulations at various temperature were performed.
The results showed that the activation energy of the Si-face is much larger than that of the C-face.
In the case of the Si-face, a flat and aligned interface structure including Si${}^{1+}$ was created.
Based on the estimated activation energies of the intermediate oxide states, it is proposed that the stability of the flat interface structure is the origin of the high activation energy of the oxidation of the Si-face.
In contrast, in the case of the C-face, it is found that the Si atom at the interface are easily pulled up by the O atoms.
This process generates the disordered interface and decreases the activation energy of the oxidation.
It is also proposed that many excess C atoms are created in the case of the C-face.

\end{abstract}


\maketitle

\section{Introduction}

Silicon carbide (SiC) is an attractive semiconductor material for applications in power electronic devices since it has excellent physical properties including wide band gap, high critical field strength and high thermal conductivity.
Among the many polytypes of SiC, 4H-SiC is considered to be particularly suitable for power electronic devices.
One of the advantages of SiC is the ability to make SiO${}_2$ insulating film by thermal oxidation as well as by using silicon, which facilitates production of the metal oxide semiconductor field effect transistor (MOSFET).

Although the SiC crystal is structually isomorphic with silicon, the thermal oxidation process of SiC is known to be very different from that of silicon.
Previous studies have reported that the SiC/SiO${}_2$ interface is neither smooth nor abrupt, \cite{:/content/aip/journal/apl/93/2/10.1063/1.2949081, hornetz1994arxps, doi:10.1116/1.580951, hornetz1994, doi:10.1063/1.371363, Imai2006547} in contrast to the Si/SiO${}_2$ interface.
The imperfection of the interface structure causes a high interface state density, which is considered to be a reason for the low channel mobility of SiC-MOSFET. \cite{doi:10.1063/1.1314293, :/content/aip/journal/apl/95/3/10.1063/1.3144272}
Therefore, a high quality SiC/SiO${}_2$ interface has not yet been developed for the fabrication of SiC devices.

The anomalousness of the oxidation mechanism of SiC appears in its kinetics.
It is known that the oxidation rate strongly depends on the surface orientation.
The rate on the ($0001$) Si-face is about one order of magnitude slower than that on the ($000\overline{1}$) C-face. \cite{Song2004}
The defect density of the SiC/SiO${}_2$ interface of the C-face is higher than that of Si-face. \cite{PSSA:PSSA321}

In a recent study, \cite{:/content/aip/journal/jap/117/9/10.1063/1.4914050} the activation energies for the oxidation were obtained by fitting the experimental data to the modified Deal-Grove formulation. \cite{massoud1985thermal}
The authors showed that the parabolic constant (corresponding to the diffusion-limiting step) was not dependent on the surface orientation and was almost equal to that of the silicon oxidation.
On the other hand, the activation energy of the linear constant (corresponding to the interface reaction-limiting step) of the Si-face was three times higher than that of C-face.

Density functional theory (DFT) calculations have been performed to investigate the atomistic process of the oxidation of SiC at the interface.
Recently, several authors have performed DFT calculations of SiC oxidation at finite temperature and reproduced the elementary processes of this oxidation, such as the dissociation of O${}_2$ molecules and the formation of CO/CO${}_2$ molecules. \cite{NIMS_SiC_oxidation63jsap, NIMS_SiC_oxidation77jsap}
However, because of its limited applicability, a long-term simulation to reproduce the continuous oxidation process including the progress of the interface has not been achieved.
For investigation of the kinetics of the oxidation of SiC, the classical molecular dynamics (MD) is a suitable method.
However, to the best of our knowledge, there is no interatomic potential to reproduce the surface orientation-dependence of the kinetics of the oxidation of SiC.

In our previous studies, \cite{TakamotoGraphene2018, Takamoto2016, KumagaiFitting} we developed a hybrid charge-transfer-type interatomic potential, based on the Tersoff potential, into which we incorporated the covalent-ionic mixed bond nature.
Using the framework of our proposed potential, we previously developed an Si-O potential for the thermal oxidation of silicon \cite{Takamoto2016} and Si-C potential for graphene growth on SiC substrate. \cite{TakamotoGraphene2018}
In this paper, we developed a new Si-O-C interatomic potential to reproduce the kinetics of the oxidation of SiC, with a particular focus on the reactions at the SiC/SiO${}_2$ interface.
Our aim in this paper was to reproduce the surface orientation-dependence of the rate of the oxidation process and to reveal its mechanism using the newly developed interatomic potential.

\section{\label{sec:Method}Methods}

In addition to the potential parameters for the already-developed Si-O and Si-C systems, parameters for C-O and Si-O-C (the three-body term) are needed.
The function form of the interatomic potential is shown in our previous papers. \cite{TakamotoGraphene2018, Takamoto2016}

We fit the potential parameters using our potential-making scheme. \cite{Takamoto2016, KumagaiFitting}
For C-O parameters, the sampled structures other than the CO and CO${}_2$ molecules used for fitting are described below.
Structures in which an O atom is connected to a C cluster (chain or ring) are employed.
To fit the energy surface of the desorption of molecules from the C cluster, MD simulations to separate CO or CO${}_2$ molecules from the edge of the C chain structure are performed to sample the snapshots.
MD simulations of the O${}_2$ molecule insertion into the diamond are performed for sampling.
Bulk diamond structures which contain several O atoms are also created.

The remaining parameters are the three-body parameters which contain each of the Si, C and O atoms.
First, amorphous Si${}_4$C${}_{4-x}$O${}_{2}$ oxycarbide structures are created.
Oxycarbide structures with excess Si, C or O atoms are also created.
For structures including molecules, SiO${}_2$ structures with CO molecules and those with CO${}_2$ molecules are created.

After fitting the above structures, additional fitting using the snapshots of DFT calculations for the SiC oxidation process \cite{NIMS_SiC_oxidation63jsap, NIMS_SiC_oxidation77jsap} is performed.
The systems include 250 atoms.
Those snapshots involves various reaction pathways such as the dissociation of O${}_2$ molecules, the bond formations of Si-O and C-O, the bond recombination, the formations of C-C bonds at the SiC/SiO${}_2$ interface and the formations of CO and CO${}_2$ molecules.
As a result of the above reactions, the following structures are also included in the snapshots: C clusters (chain and ring) at the interface, Si atoms with intermediate oxide states such as Si${}^{1+}$, Si${}^{2+}$ and Si${}^{3+}$ and molecules (O${}_2$, CO and CO${}_2$) in the amorphous SiO${}_2$.
The number of material properties used for fitting reaches about 1,000,000.
It is noted that parameters for Si-O are refitted in this paper because the potential functions were changed from the previous paper. \cite{Takamoto2016}

\begin{figure}[tbp]
	\centering
	\begin{minipage}[t]{.49\linewidth}
		\centering
		\includegraphics[width=1.\linewidth,trim=40 00 40 00]{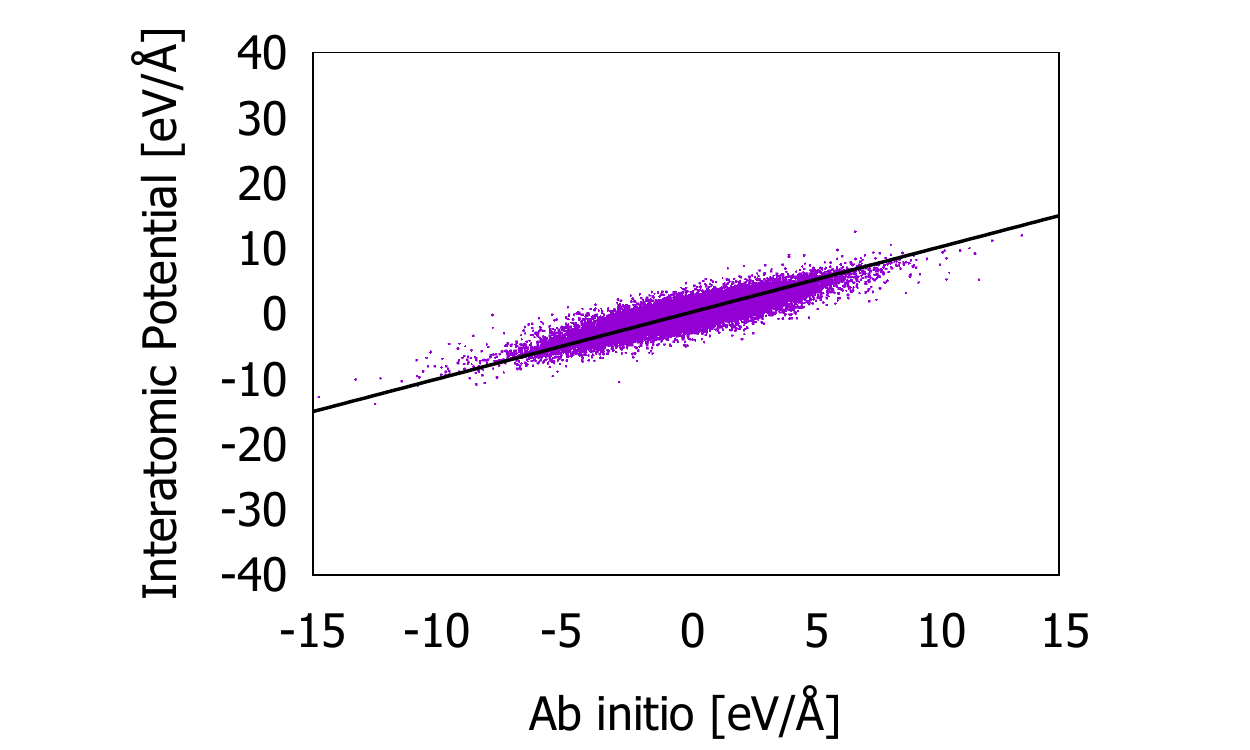}
		\subcaption{This work}
		\label{fig:force_sico_thiswork}
	\end{minipage}
	\begin{minipage}[t]{.49\linewidth}
		\centering
		\includegraphics[width=1.\linewidth,trim=40 00 40 00]{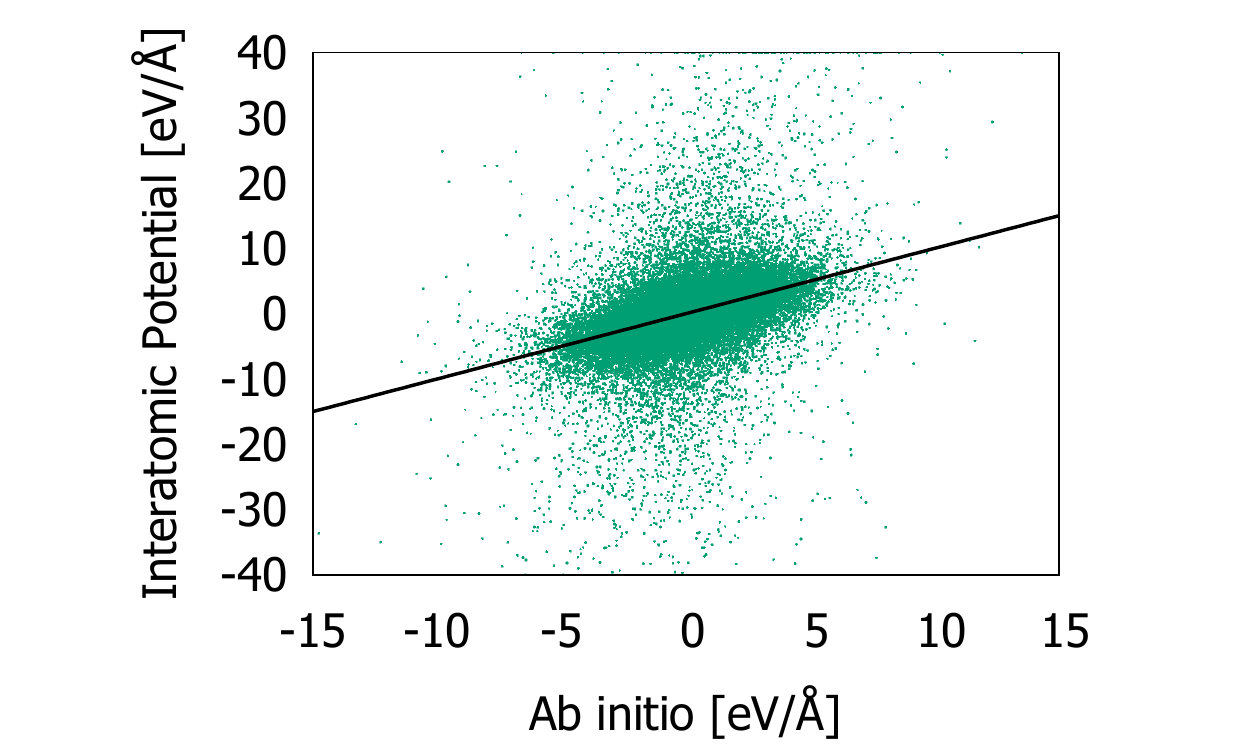}
		\subcaption{ReaxFF  \cite{doi:10.1021/jp306391p}}
		\label{fig:force_sico_withreaxff}
	\end{minipage}
	\caption{Force comparison between Ab initio calculation and interatomic potentials.}
	\label{fig:force_sico}
\end{figure}

Figure \ref{fig:force_sico}(\subref{fig:force_sico_thiswork}) plots the comparison of the atomic forces between DFT calculations and the developed interatomic potential.
The case for ReaxFF \cite{doi:10.1021/jp306391p} is also shown in \figref{fig:force_sico}(\subref{fig:force_sico_withreaxff}).
The correlation coefficient of our interatomic potential is 0.91, while that of ReaxFF is 0.43.
It is proved that our interatomic potential well reproduces the various structures.
This interatomic potential is implemented in the LAMMPS Molecular Dynamics Simulator. \cite{plimpton1995fast, lammpssource}
In our simulations, the calculation speed of our interatomic potential is slightly faster than that of ReaxFF.

\section{\label{sec:Sim}SiC oxidation simulation}

\begin{figure}[tbp]
	\centering
	\begin{minipage}[t]{.49\linewidth}
		\centering
		\includegraphics[width=1.\linewidth,trim=00 00 00 00]{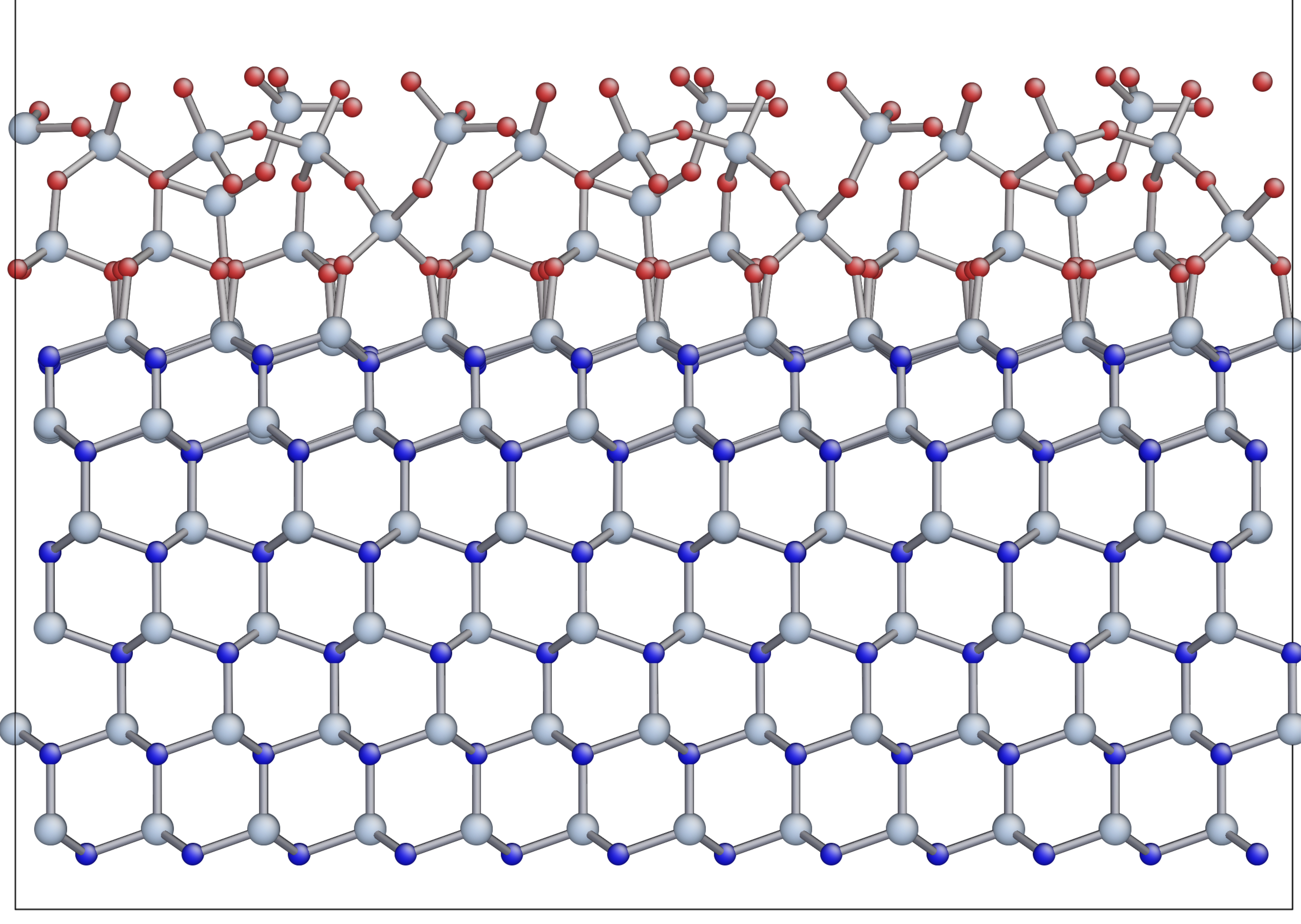}
		\subcaption{Si-face}
	\end{minipage}
	\begin{minipage}[t]{.49\linewidth}
		\centering
		\includegraphics[width=1.\linewidth,trim=00 00 00 00]{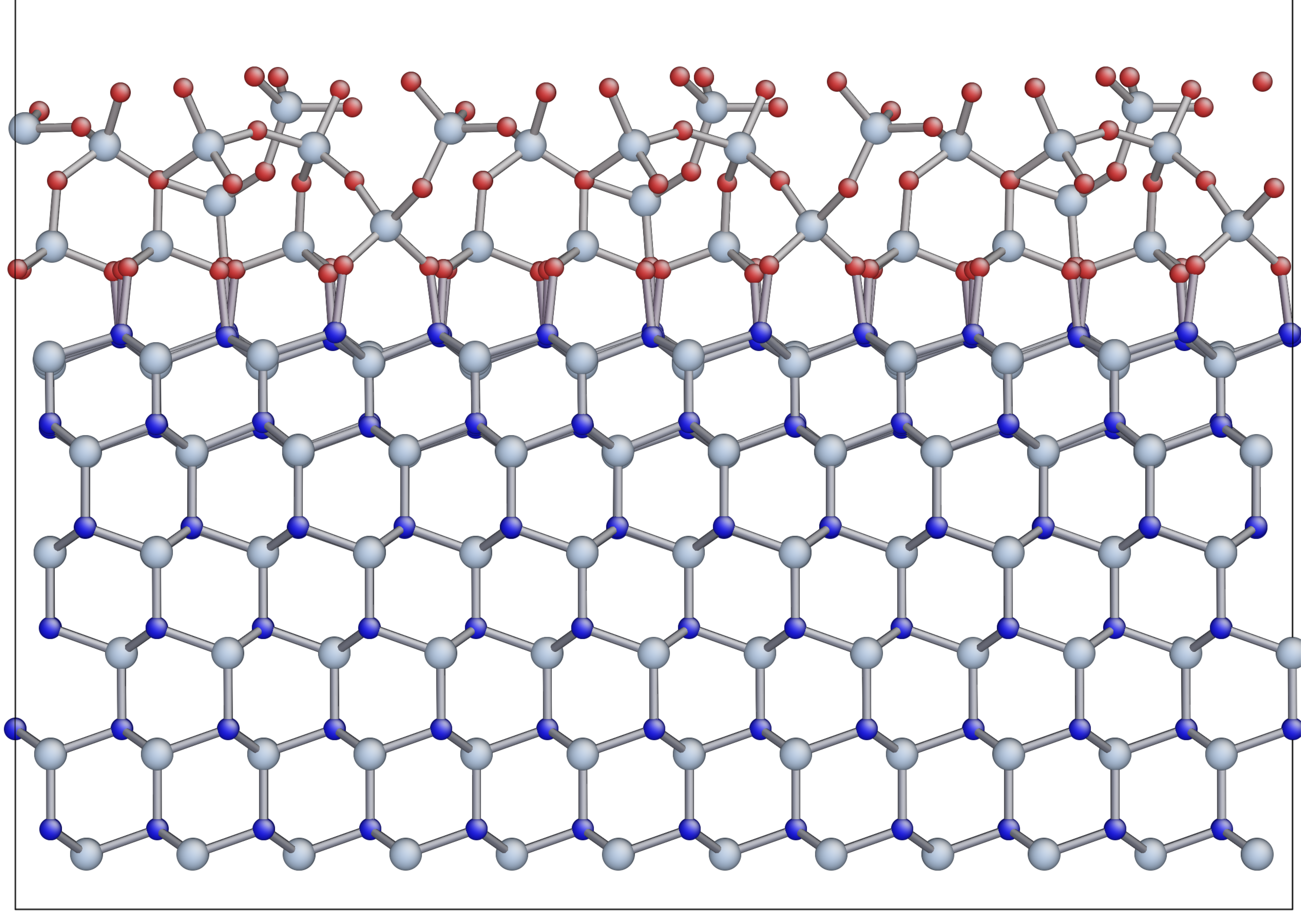}
		\subcaption{C-face}
	\end{minipage}
	\caption{Initial structure for SiC oxidation simulation. The orientation perpendicular to the paper corresponds to ($11\overline{2}0$). Gray spheres are Si atoms, red spheres are O atoms and blue spheres are C atoms.}
	\label{fig:sic_md_init}
\end{figure}

The oxidation simulation of SiC is performed with the developed interatomic potential.
The initial structure is shown in \figref{fig:sic_md_init}.
A thin SiO${}_2$ layer is connected to the SiC surface.
The area of the interface is about $30\times 25$ $\mathrm{nm}^2$.
The structure of the C-face is also created by replacing Si atoms with C atoms and vice versa.

In order to focus on the reaction of O${}_2$ molecules, the oxidation simulation is realized by inserting O${}_2$ molecules into SiO${}_2$ repeatedly.
The details of the simulation are as follows:
\begin{enumerate}
\item The initial SiO${}_2$ layer is annealed for 2 ps (5000 K for 1 ps and 2000 K for 1 ps) while fixing the other regions.
\item Then, the O${}_2$ molecules are inserted into the SiO${}_2$ region. The number of O${}_2$ molecules to be inserted is set so that its density becomes about 1 molecule/nm${}^3$. The inserted position is randomly selected from the points that are more than 1.8 $\mathrm{\AA}$ apart from all other atoms. \label{enum:loop}
\item 1 ps of MD (NVT ensemble) at certain temperature with fixed lattice size taking into account the thermal expansion is performed while fixing the bottom bilayer of SiC.
\item After 1 ps, the remaining O${}_2$ molecules are all removed. All CO and CO${}_2$ molecules generated during the simulation are removed every 1 ps, since those diffusions are fast enough to ignore, unlike O${}_2$ molecules. \cite{:/content/aip/journal/jap/117/9/10.1063/1.4914050}
\item Go back to \ref{enum:loop} and iterate. Insertion of O${}_2$ molecules and 1 ps of MD are performed in the same manner.
\end{enumerate}

These insertion procedures enable us to boost the diffusion of O${}_2$ molecules in SiO${}_2$.
The timestep is set to 0.5 fs. The periodic boundary condition is applied.

\begin{figure}[tbp]
	\begin{minipage}[t]{.49\linewidth}
		\centering
		\includegraphics[width=1.\linewidth,trim=00 00 00 00]{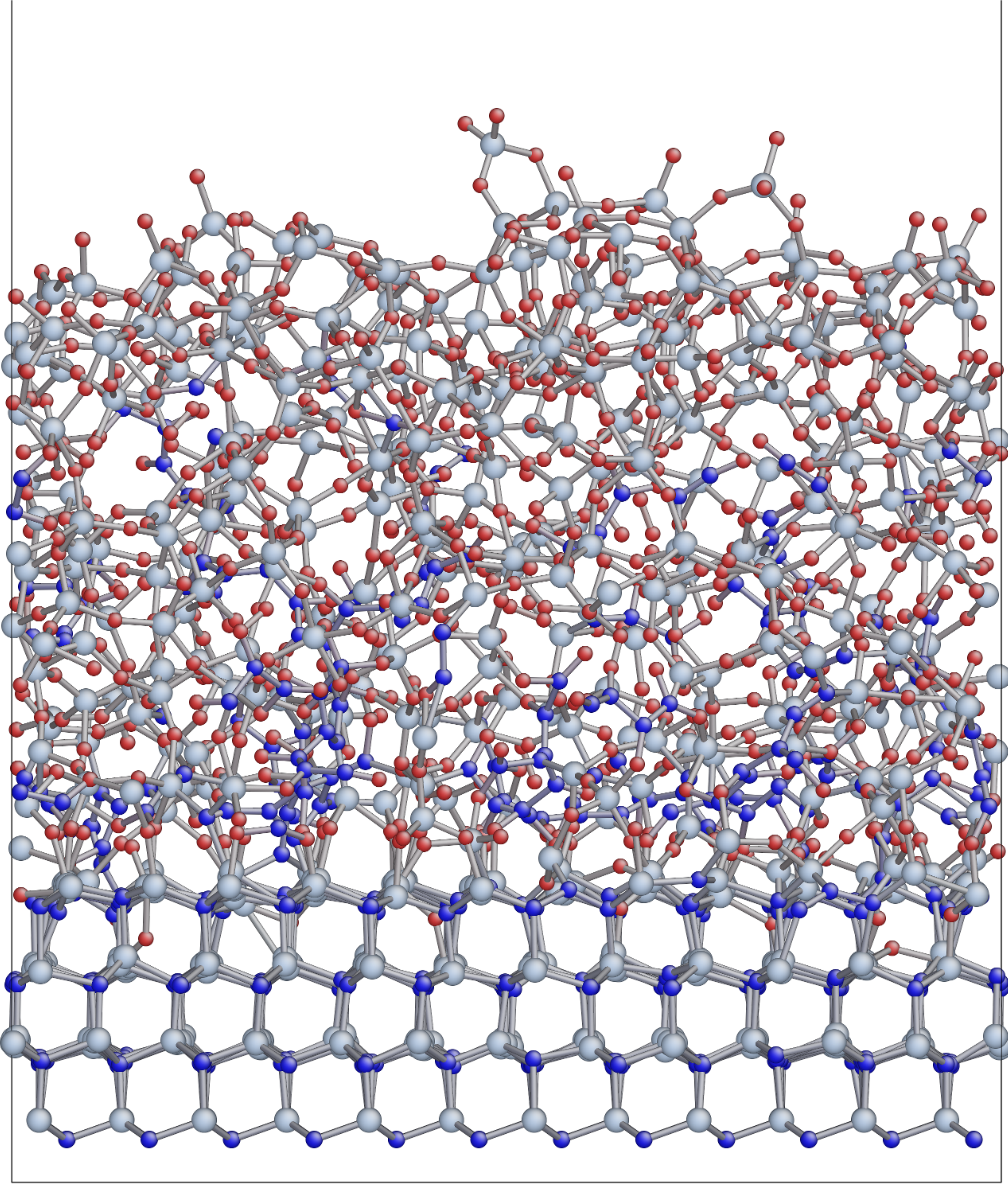}
		\subcaption{Si-face after 140 ps}
	\end{minipage}
	\begin{minipage}[t]{.49\linewidth}
		\centering
		\includegraphics[width=1.\linewidth,trim=00 00 00 00]{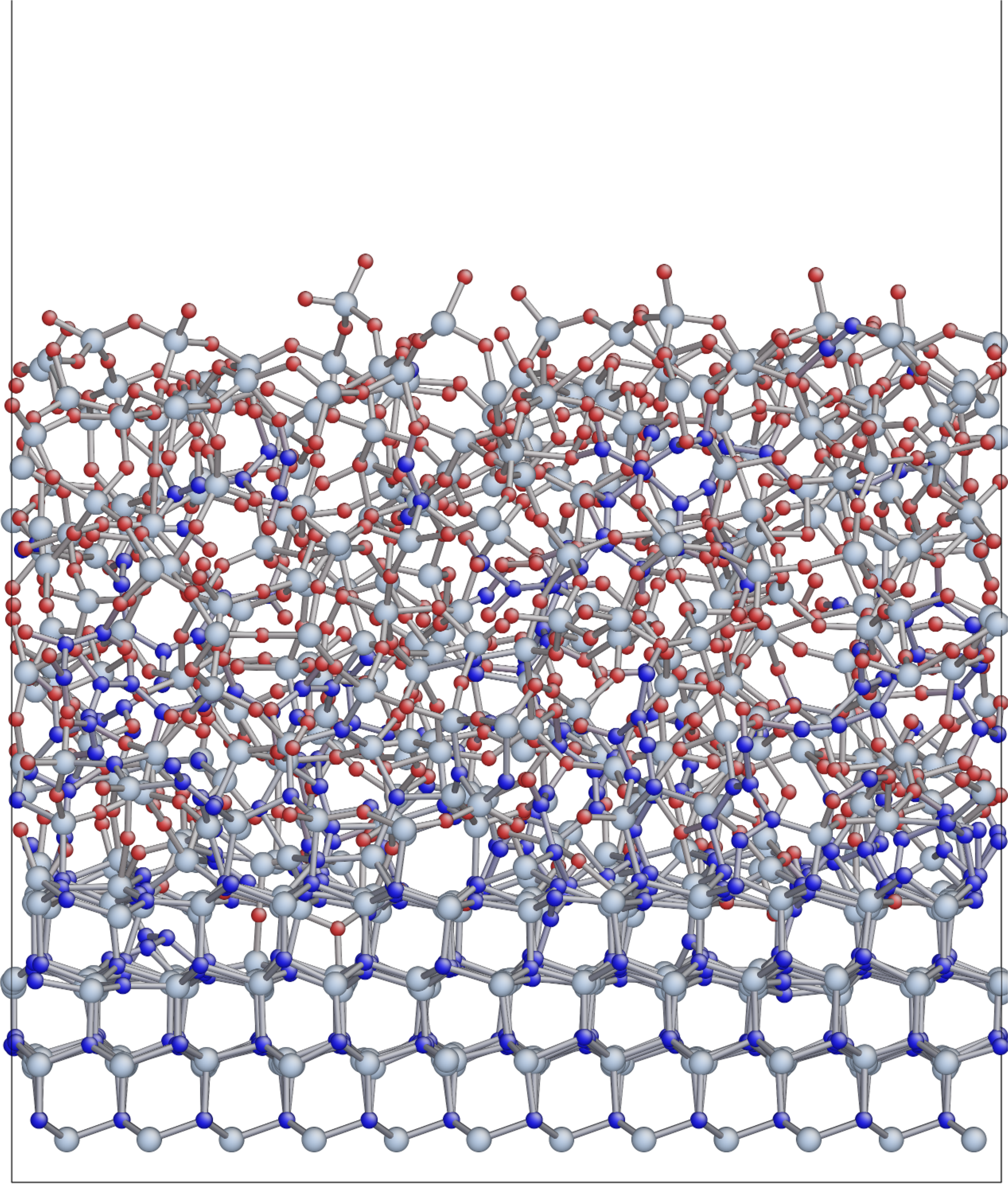}
		\subcaption{C-face after 70 ps}
	\end{minipage}
	\caption{Comparison of the interface structures during oxidation simulation. Gray spheres are Si atoms, red spheres are O atoms and blue spheres are C atoms.}
	\label{fig:sic_md_oxidation_snapshots}
\end{figure}

\begin{figure}[tbp]
	\begin{minipage}[t]{.49\linewidth}
		\centering
		\includegraphics[width=1.\linewidth,trim=00 00 00 00]{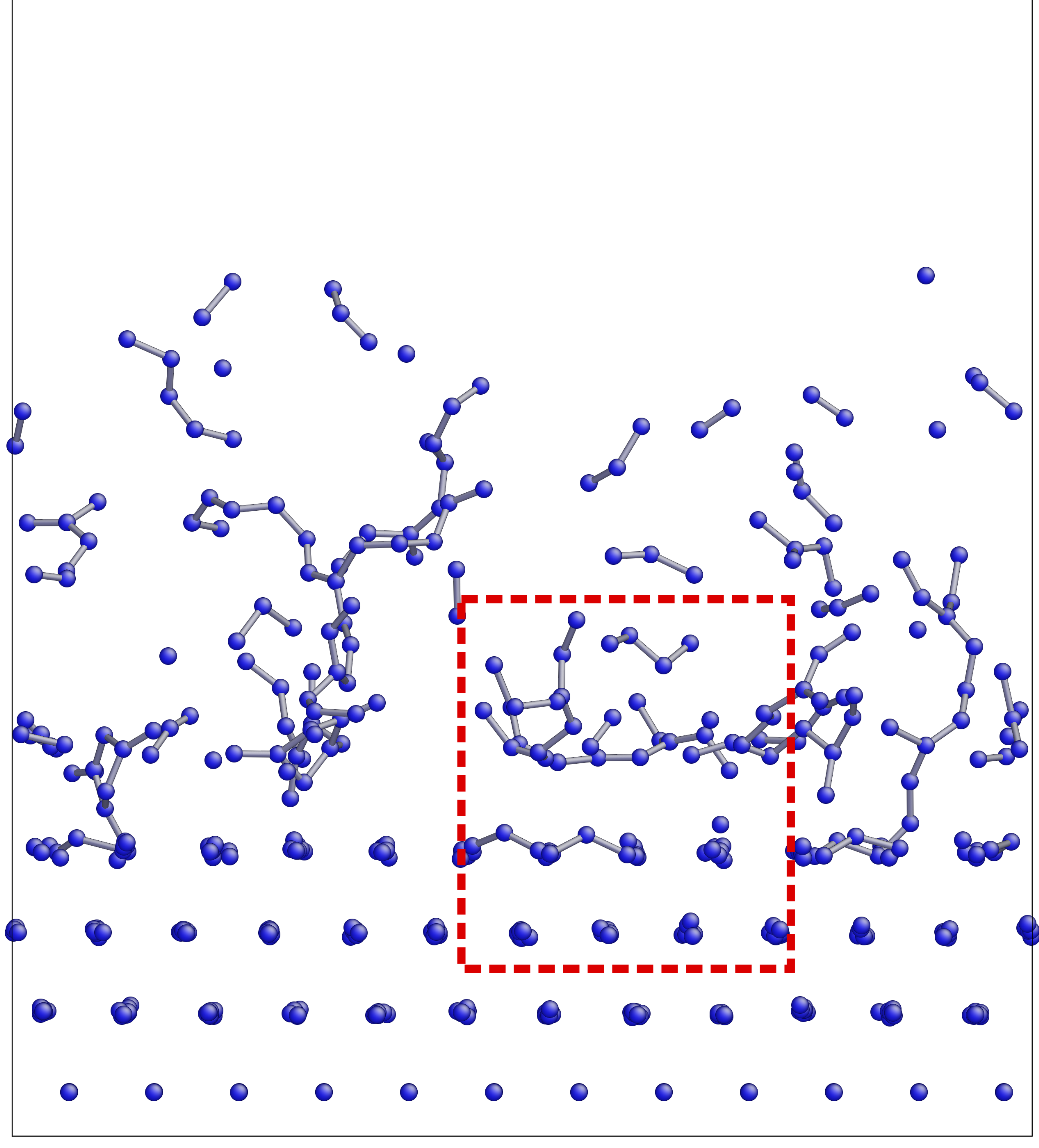}
		\subcaption{Si-face after 140 ps}
	\end{minipage}
	\begin{minipage}[t]{.49\linewidth}
		\centering
		\includegraphics[width=1.\linewidth,trim=00 00 00 00]{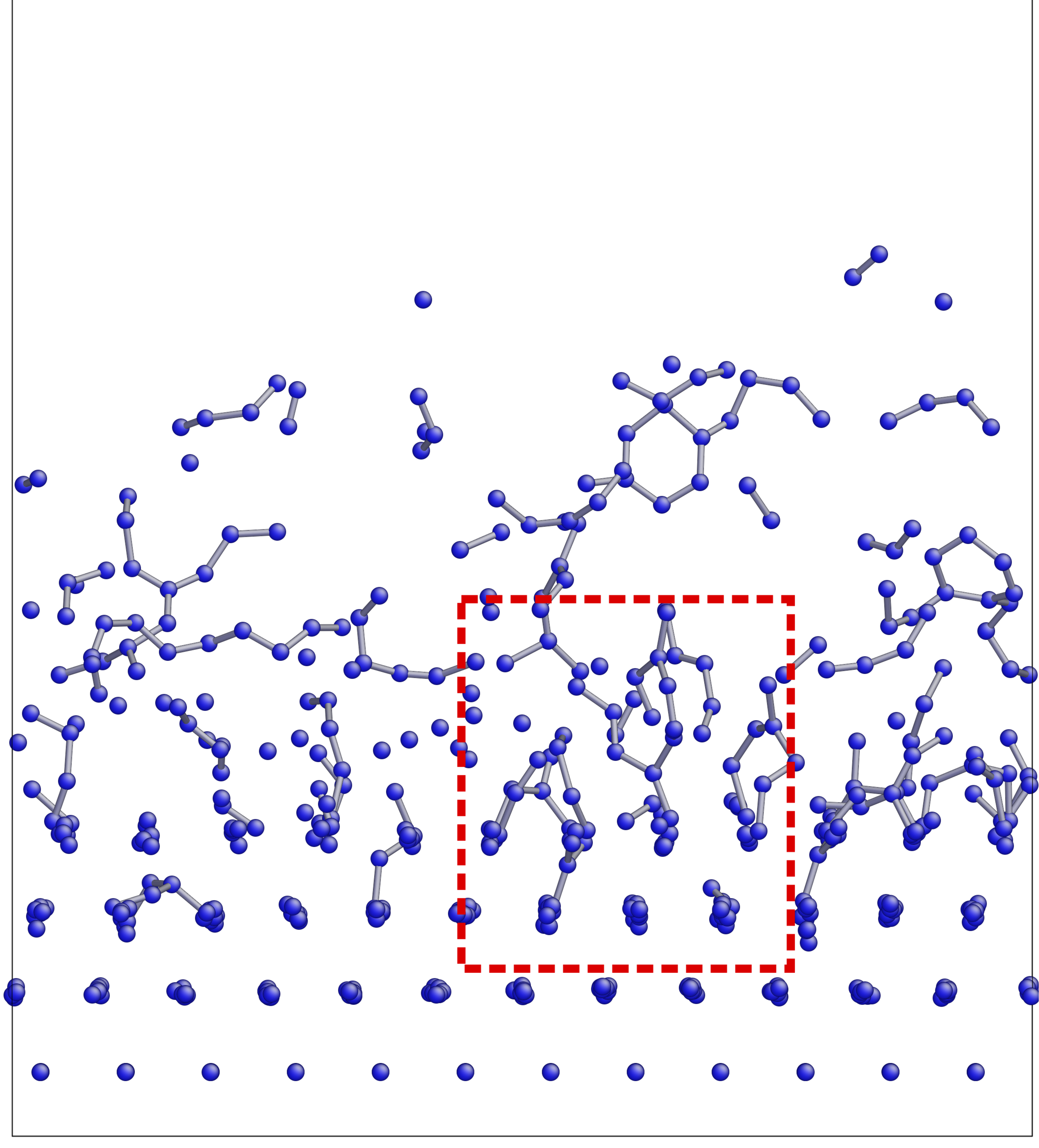}
		\subcaption{C-face after 70 ps}
	\end{minipage}
	\caption{Comparison of the interface structures during oxidation simulation (C only).}
	\label{fig:sic_md_oxidation_snapshots_c}
\end{figure}

In the case of 1600 K oxidation, the progress in oxidation is observed in both the Si-face and C-face cases.
Figure \ref{fig:sic_md_oxidation_snapshots} shows the snapshots of the oxidation simulation of both faces, in which the same amount of SiC is oxidized.
The oxidation rate of the C-face is faster than that of the Si-face.
Figure \ref{fig:sic_md_oxidation_snapshots_c} shows only the C atoms for the same snapshots.

\begin{figure}[tbp]
	\begin{minipage}[t]{1.\linewidth}
		\centering
		\includegraphics[width=1.\linewidth,trim=00 00 00 00]{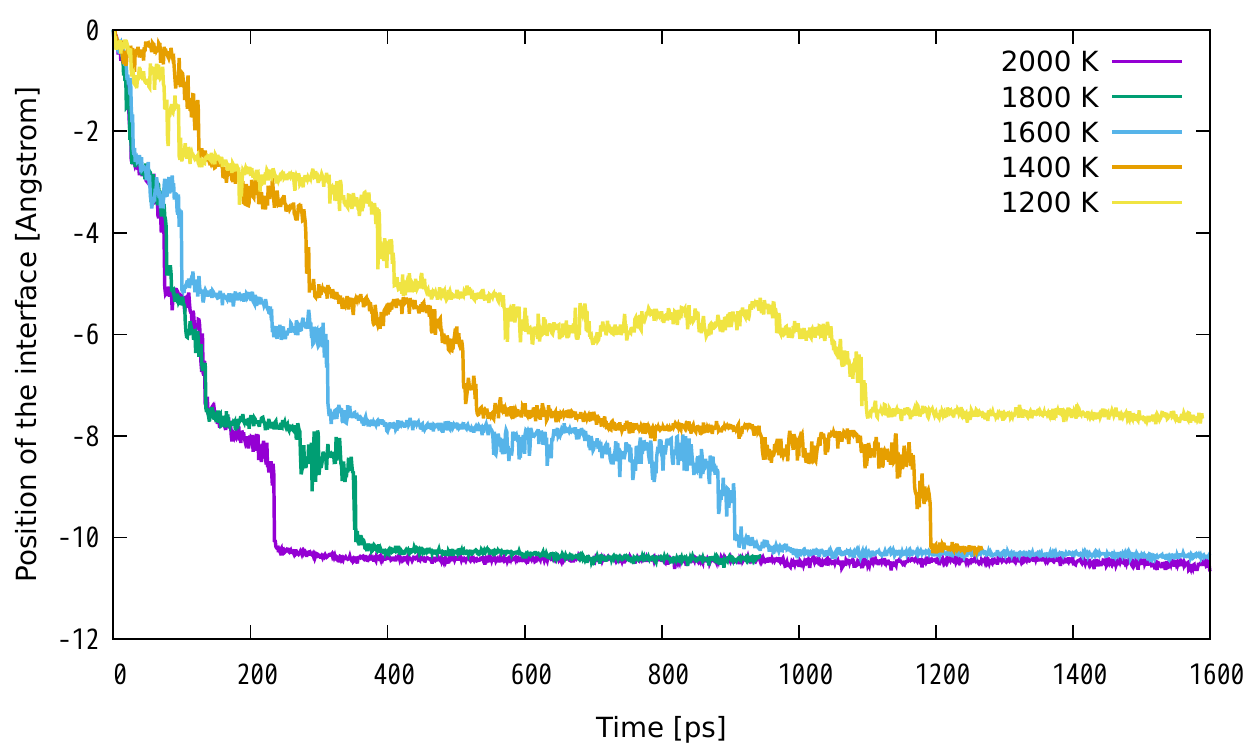}
		\subcaption{Si-face}
	\end{minipage}
	\begin{minipage}[t]{1.\linewidth}
		\centering
		\includegraphics[width=1.\linewidth,trim=00 00 00 00]{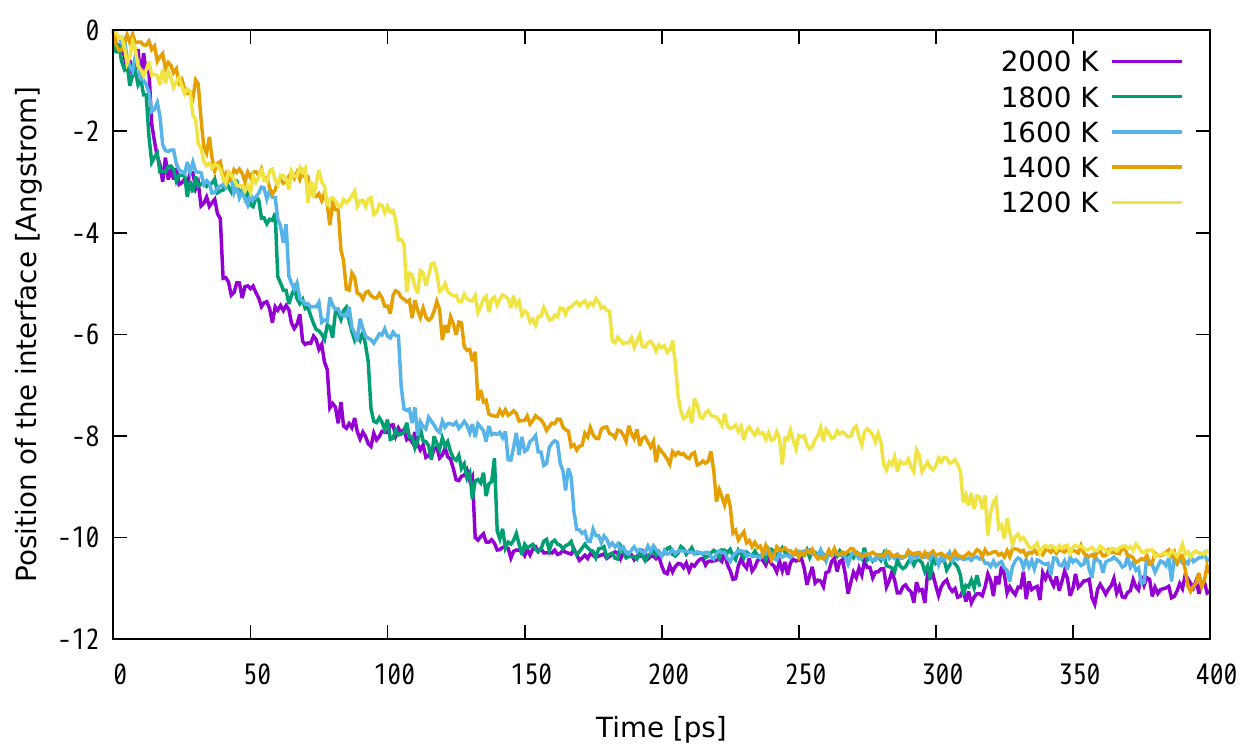}
		\subcaption{C-face}
	\end{minipage}
	\caption{Temperature-dependence of the rate of progress of the oxidation.}
	\label{fig:sic_md_oxidation_temp}
\end{figure}

\begin{figure}[tbp]
	\centering
	\includegraphics[width=.89\linewidth,trim=00 00 00 00]{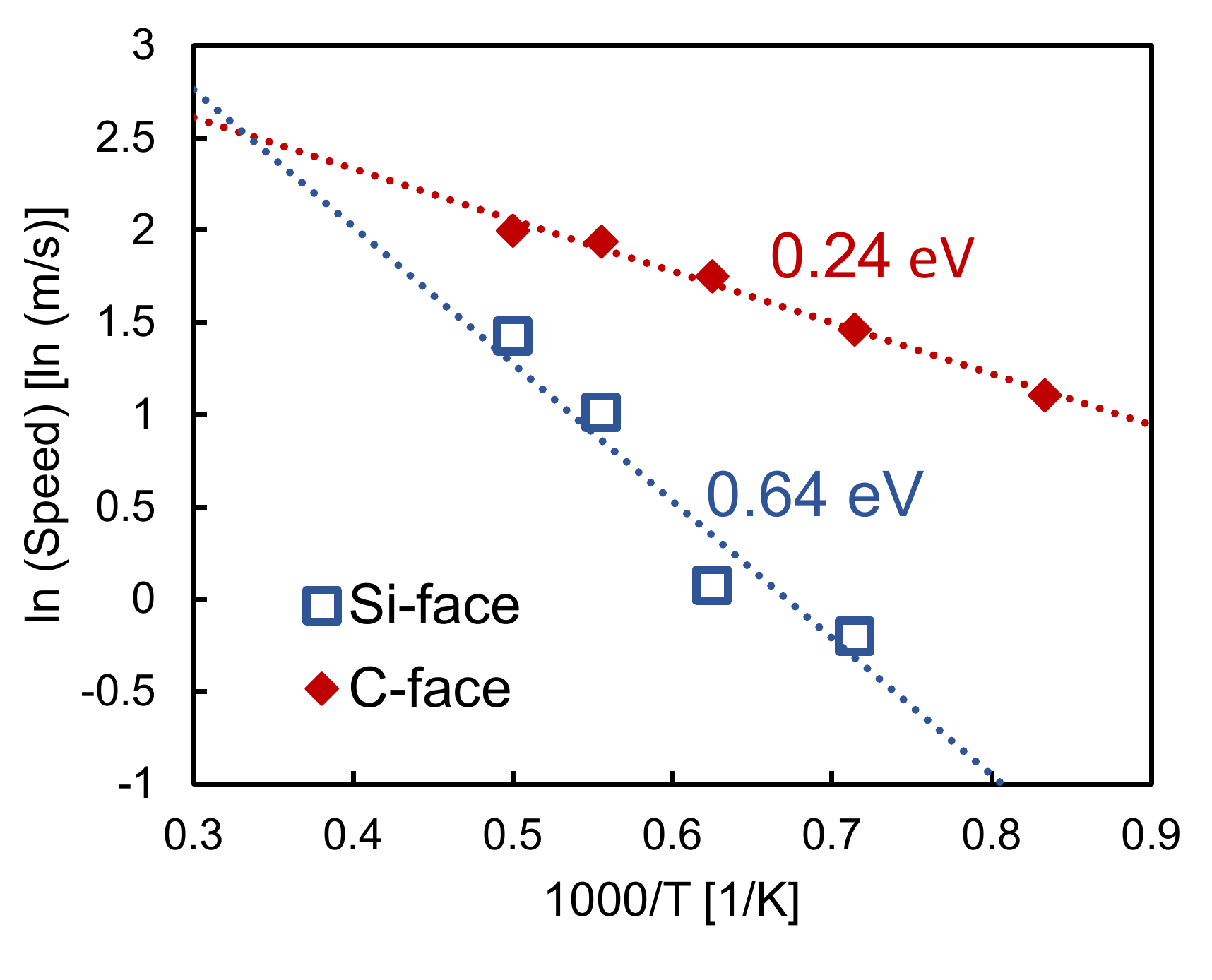}
	\caption{Arrhenius plot of the rate of progress of oxidation.}
	\label{fig:sic_md_oxidation_arrhenius}
\end{figure}

Next, a series of oxidation simulations are performed by changing the temperature from 1200 K to 2000 K in increments of 200 K.
Figure \ref{fig:sic_md_oxidation_temp} shows the time histories of the positions of the SiC/SiO${}_2$ interface.
It is noted that the fixed bottom layer is at $-12.5 \mathrm{\AA}$.
The averaged oxidation rate is calculated after the interface reaches $-9.8 \mathrm{\AA}$.
The Arrhenius plot of the oxidation rate is shown in \figref{fig:sic_md_oxidation_arrhenius}.
It indicates that the activation energy of the Si-face is 0.64 eV, which is 2.7 times larger than that of the C-face (0.24 eV).
Previous experimental work \cite{:/content/aip/journal/jap/117/9/10.1063/1.4914050} reported that the activation energy of the linear rate constant of the Deal-Grove model for the Si-face is 2.9 times larger than that of the C-face, which is consistent with our results.

It is noted that the quantitative values of the activation energy are smaller than those of the experimental data.
This might be due to the difference in oxide thickness.
It is well known that the growth rate is higher when the oxide is thin. \cite{1347-4065-47-10R-7803, 1347-4065-46-8L-L770}
In the case of the oxidation of silicon, the experimental data of the activation energy for ultrathin oxide \cite{Watanabe1998} is lower than that for thick oxide . \cite{Deal1965, Massoud1985}
DFT calculations \cite{Akiyama2005L65, PhysRevLetters93.086102} show that the activation energy for the dissociation of O${}_2$ molecules is consistent with that of ultrathin oxide.
It is considered that the effects related to thick oxide, such as roughness or stress accumulation, increase the activation energy. \cite{Akiyama2005L65}

\section{\label{sec:Discussion}Discussion}

\begin{figure}[tbp]
	\begin{minipage}[t]{.49\linewidth}
		\centering
		\includegraphics[width=1.\linewidth,trim=60 00 60 00]{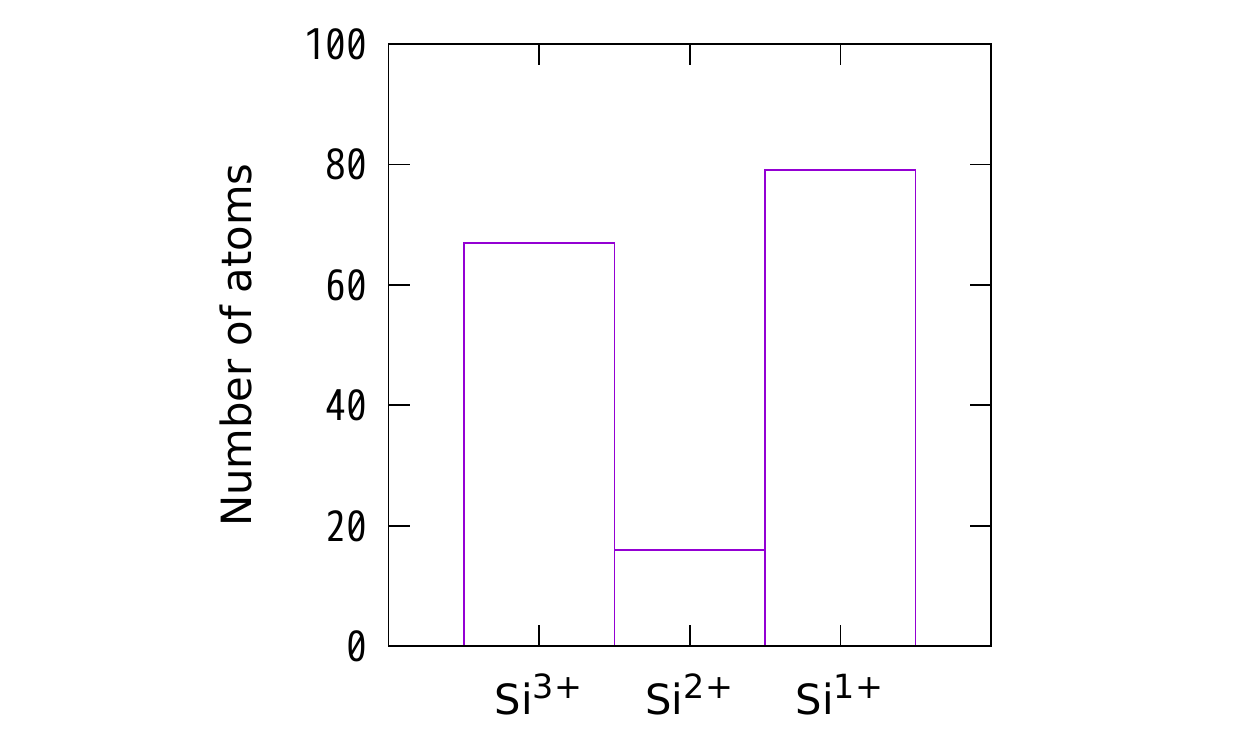}
		\subcaption{Si-face}
		\label{fig:suboxide_bar_si}
	\end{minipage}
	\begin{minipage}[t]{.49\linewidth}
		\centering
		\includegraphics[width=1.\linewidth,trim=60 00 60 00]{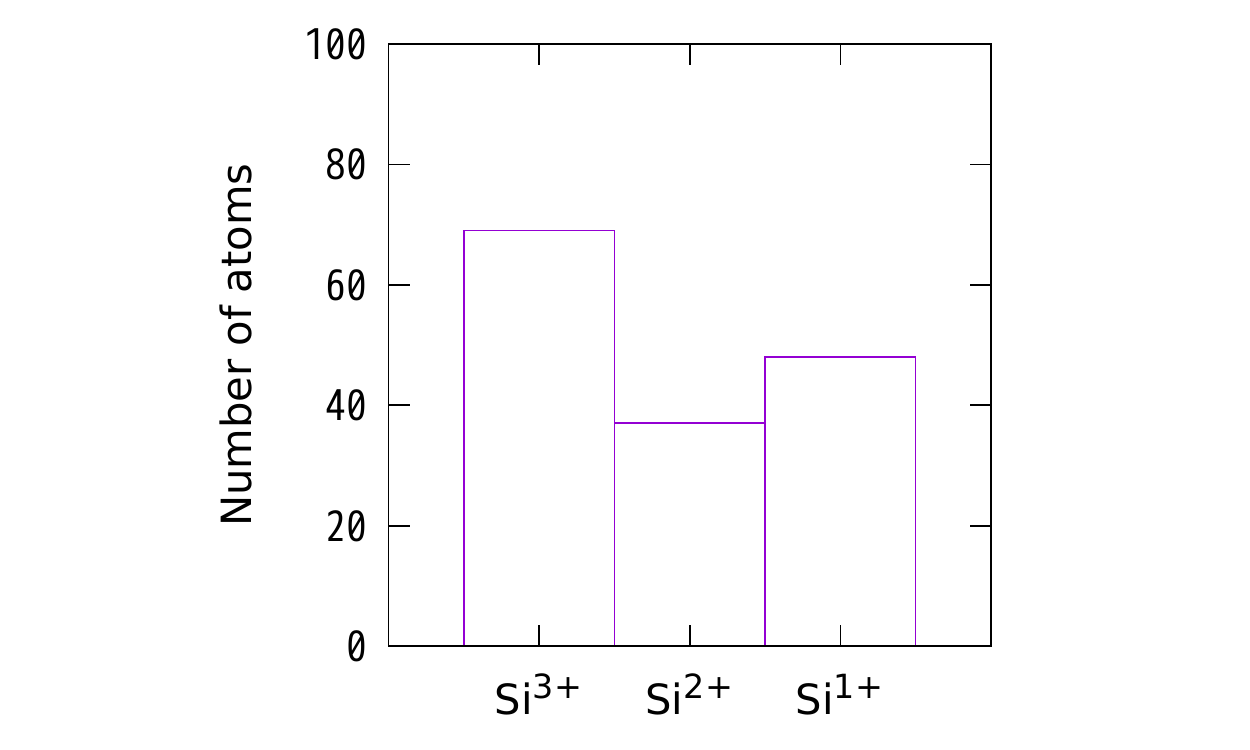}
		\subcaption{C-face}
		\label{fig:suboxide_bar_c}
	\end{minipage}
	\caption{Comparison of numbers of Si${}^{1+}$, Si${}^{2+}$ and Si${}^{3+}$.}
	\label{fig:sic_md_oxidation_suboxide_bar}
\end{figure}

In order to investigate the mechanism underlying the dependence of the oxidation rate on the crystal orientation of SiC, we focus on the valences of the Si atoms at the interface.
In the case of the Si-face (\figref{fig:sic_md_oxidation_suboxide_bar}(\subref{fig:suboxide_bar_si})), the number of Si${}^{1+}$ is larger than that of Si${}^{3+}$, in contrast to the case of the C-face (\figref{fig:sic_md_oxidation_suboxide_bar}(\subref{fig:suboxide_bar_c})).
Also, the number of Si${}^{2+}$ is smaller than that of the C-face.
These results are in good agreement with the XPS observation. \cite{watanabe2012synchrotron, :/content/aip/journal/apl/99/2/10.1063/1.3610487}

\begin{figure}[tbp]
	\centering
	\includegraphics[width=1.\linewidth,trim=00 00 00 00]{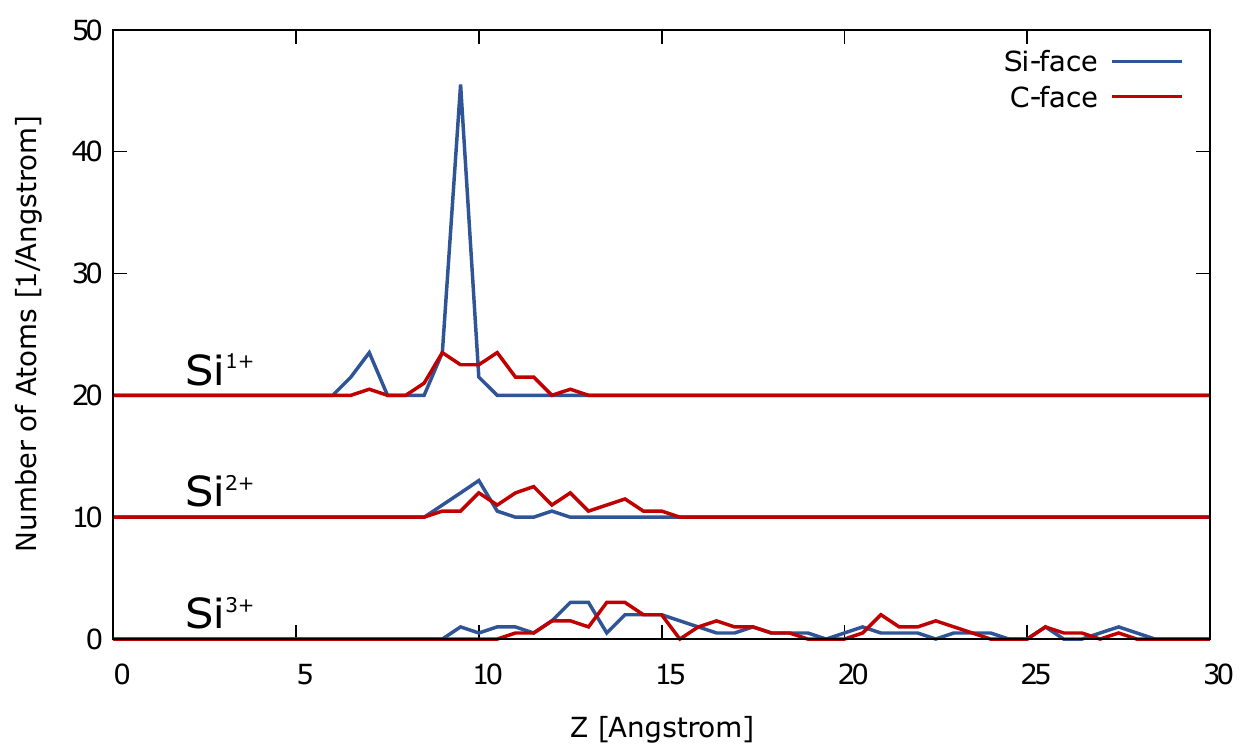}
	\caption{Distribution of Si${}^{1+}$, Si${}^{2+}$ and Si${}^{3+}$ obtained by the oxidation simulation. Orientation of the z axis is perpendicular to surface.}
	\label{fig:sic_md_oxidation_si_hist}
\end{figure}

Figure \ref{fig:sic_md_oxidation_si_hist} shows the distribution of Si${}^{1+}$, Si${}^{2+}$ and Si${}^{3+}$ along the direction perpendicular to the surface.
Only in the case of the Si-face, the distribution of Si${}^{1+}$ has a sharp peak.
Those Si atoms are located on the interface front of the SiC substrate and have one vertical Si-O bond and three Si-C back bonds, as seen in \figref{fig:sic_md_oxidation_snapshots}.

The lifetimes of Si${}^{1+}$, Si${}^{2+}$ and Si${}^{3+}$ are also investigated.
Assuming that Si${}^{n+}$ stochastically changes to Si${}^{(n+1)+}$ ($n=1$ or $2$ or $3$), the survival rate $P_n\left(t\right)$, the percentage of Si${}^{n+}$ which are alive for a time $t$ after their generation, can be written as

\begin{equation}
	P_n\left(t\right)=\exp{(-\lambda_n t)}, \label{eq:n_reduce}
\end{equation}
where $\lambda_n$ is the reaction rate constant.
Assuming $\lambda_n$ follows the Arrhenius equation, $\lambda_n$ can be written as

\begin{equation}
	\lambda_n\left(T\right)=A_n\exp{\left(-\frac{E_n}{k_BT}\right)},
\end{equation}
where $A_n$, $E_n$, $k_B$ and $T$ are the pre-exponential factor, the activation energy for the reaction, the Boltzmann constant and the temperature, respectively.
To calculate the activation energies $E_n$, the rate constant $\lambda_n\left(T\right)$ for various temperatures $T$ are obtained by fitting the simulation results to \eref{eq:n_reduce}.
Then, the activation energies $E_n$ for each $n$ are calculated both in the case of the Si-face and the C-face.
The results are shown in Table \ref{tab:suboxide_activation_energy}.

\begin{table}
	\caption{Calculated activation energies $E_n$.}
	\label{tab:suboxide_activation_energy}
	\begin{tabular}{lccc}
		 & \multicolumn{1}{c}{Si-face [eV]} & \multicolumn{1}{c}{C-face [eV]} & \multicolumn{1}{c}{\shortstack{Corresponding\\reaction path}} \\ \hline
		$E_1$ & $0.7$ & $0.2$ & Si${}^{1+}$ $\rightarrow$ Si${}^{2+}$ \\
		$E_2$ & $0.3$ & $0.2$ & Si${}^{2+}$ $\rightarrow$ Si${}^{3+}$ \\
		$E_3$ & $0.1$ & $0.0$ & Si${}^{3+}$ $\rightarrow$ Si${}^{4+}$
	\end{tabular}
\end{table}
$E_1$ (corresponding to the reaction of Si${}^{1+}$ $\rightarrow$ Si${}^{2+}$) in the case of the Si-face is 0.7 eV, which is much higher than the others.
Also, this value is close to the activation energy of the Si-face oxidation shown in \figref{fig:sic_md_oxidation_arrhenius} (0.64 eV).
Therefore, the reaction of Si${}^{1+}$ $\rightarrow$ Si${}^{2+}$ would be the rate-limiting process in the whole oxidation process in the case of Si-face.
On the other hand, all $E_n$ values in the case of the C-face are relatively low.

We also estimate the activation energies of the dissociation of O${}_2$ molecules at the interface by using nudged elastic band (NEB) calculations.
The activation energies are found to be in the range of 0.0-0.2 eV on both the Si-face and C-face.
It is noted that these low activation energies are close to that of the oxidation of silicon calculated by DFT. \cite{Akiyama2005L65, PhysRevLetters93.086102}

\begin{figure}[tbp]
	\centering
	\begin{minipage}[t]{1.\linewidth}
		\begin{minipage}[t]{.39\linewidth}
			\centering
			\includegraphics[width=1.\linewidth,trim=00 00 00 00]{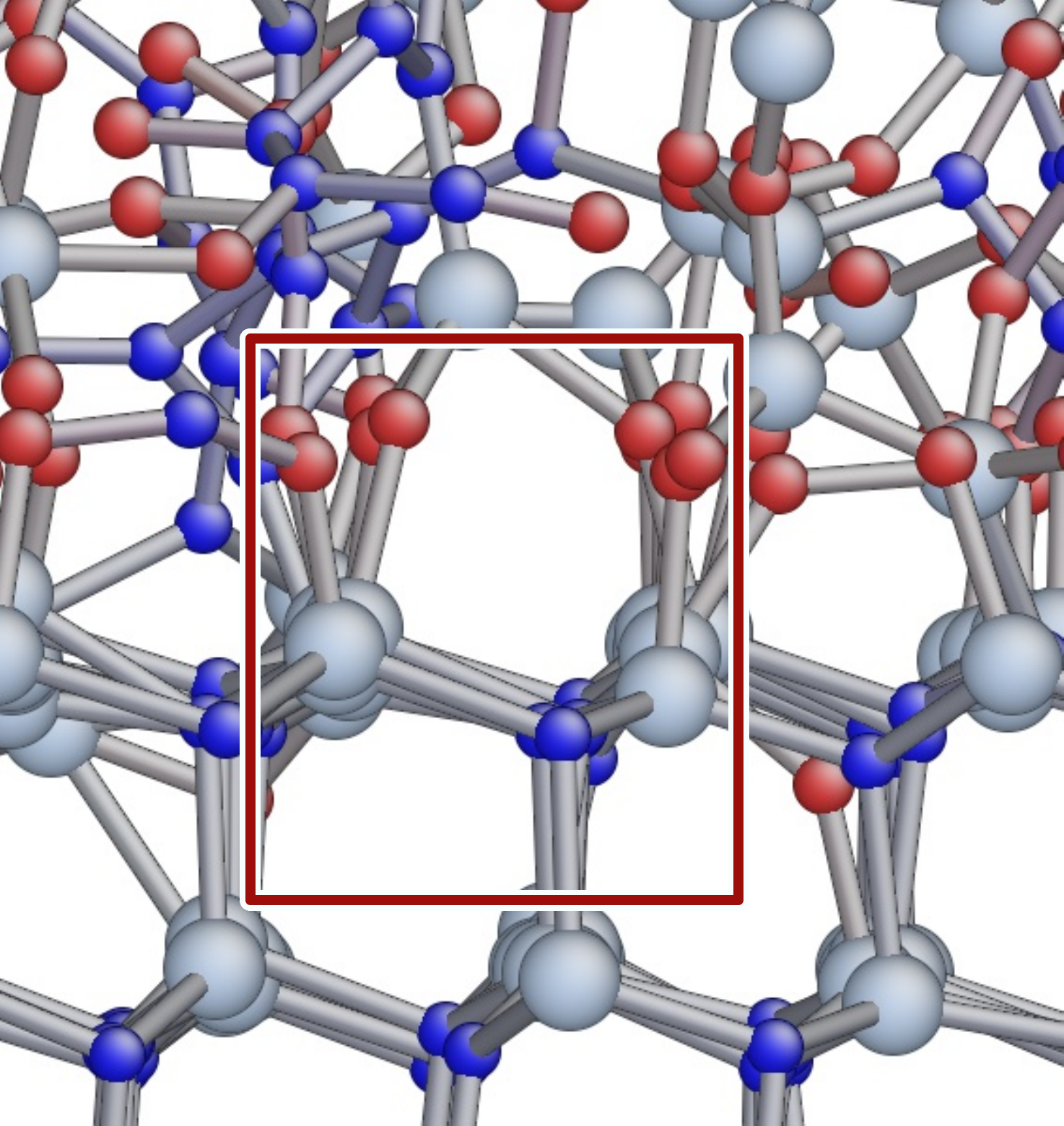}
			\subcaption{}
			\label{fig:interface_snapshot_si}
		\end{minipage}
		\begin{minipage}[t]{.59\linewidth}
			\centering
			\includegraphics[width=.5\linewidth,trim=00 00 00 00]{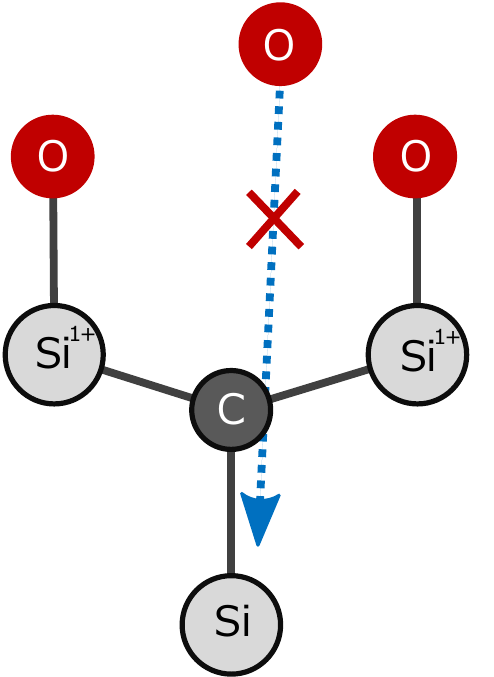}
			\subcaption{}
			\label{fig:interface_model_si}
		\end{minipage}
	\end{minipage}
	\begin{minipage}[t]{1.\linewidth}
		\begin{minipage}[t]{.39\linewidth}
			\includegraphics[width=1.\linewidth,trim=00 00 00 00]{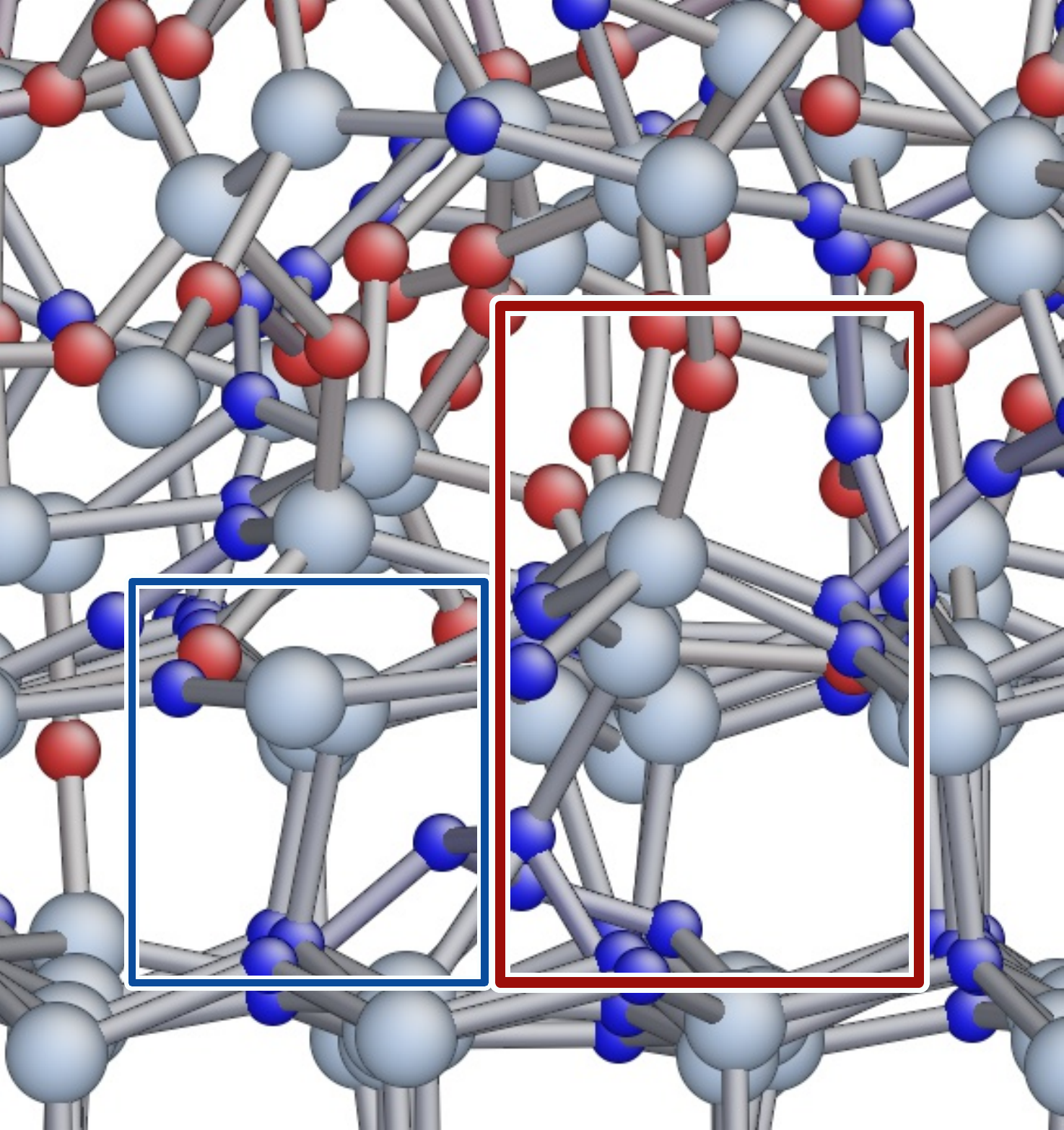}
			\subcaption{}
			\label{fig:interface_snapshot_c}
		\end{minipage}
		\begin{minipage}[t]{.59\linewidth}
			\includegraphics[width=1.\linewidth,trim=00 00 00 00]{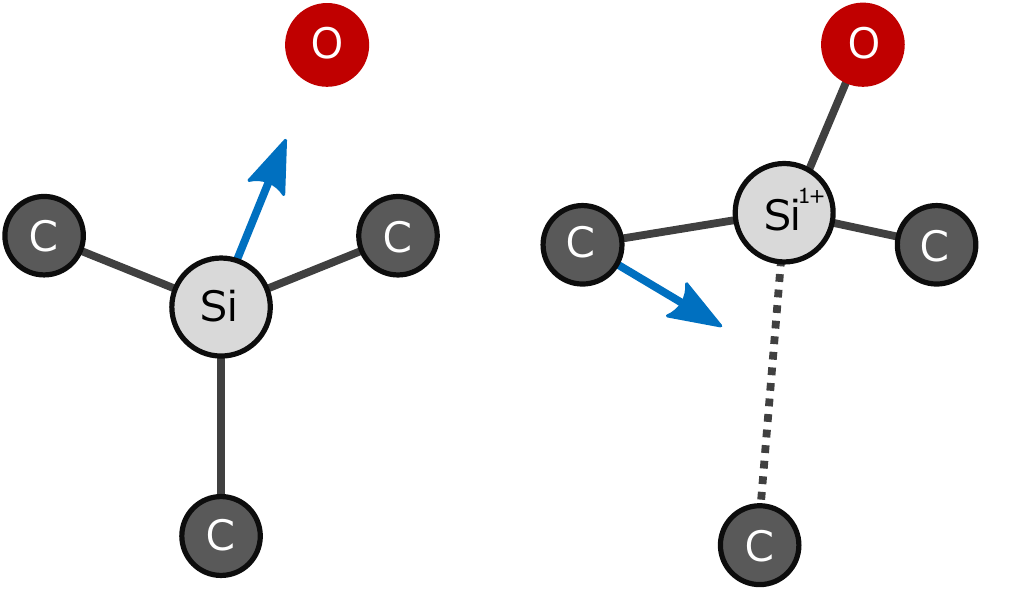}
			\subcaption{}
			\label{fig:interface_model_c}
		\end{minipage}
	\end{minipage}
	\caption{Enlarged snapshots and schematic diagrams of the SiC/SiO${}_2$ interface. (\subref{fig:interface_snapshot_si}, \subref{fig:interface_model_si}): Si-face. (\subref{fig:interface_snapshot_c}, \subref{fig:interface_model_c}): C-face.}
	\label{fig:sic_md_oxidation_model_comparison}
\end{figure}

The snapshots of the interface in the cases of Si-face (\figref{fig:sic_md_oxidation_model_comparison}(\subref{fig:interface_snapshot_si})) and C-face (\figref{fig:sic_md_oxidation_model_comparison}(\subref{fig:interface_snapshot_c})) are shown.
In the case of the Si-face, the Si atoms have one vertical front bond and three back bonds.
The flat interface can be fabricated by replacing the front Si-C bond with an Si-O bond, as is schematically shown in \figref{fig:sic_md_oxidation_model_comparison}(\subref{fig:interface_model_si}).
The Si atoms correspond to Si${}^{1+}$, which is flatly distributed at the interface, as discussed above.
For further oxidation reaction, O atoms need to access the Si atoms located in the next layer.
Since the Si atom is apart from the interface, the reaction with O atoms is inhibited.
In order for the oxidation to proceed, the aligned Si-C bonds at the interface must be broken.
That reaction would be the rate-limiting process of the oxidation (corresponding to $E_1$ shown in Table \ref{tab:suboxide_activation_energy}).

In contrast, the Si atoms in the C-face have three front bonds and one vertical back bond.
As the O atom approaches the interface, it easily pulls up the Si atom because it prefers to connect with the Si atom, as is schematically shown in \figref{fig:sic_md_oxidation_model_comparison}(\subref{fig:interface_model_c}).
Several of these reactions can be seen at the interface in \figref{fig:sic_md_oxidation_snapshots}.
Together, these processes form the disordered interface through many bond reconstructions.
Since there is no aligned structure in the case of the C-face, the oxidation can proceed without the high activation energy.

\begin{figure}[tbp]
	\includegraphics[width=.7\linewidth,trim=60 00 60 00]{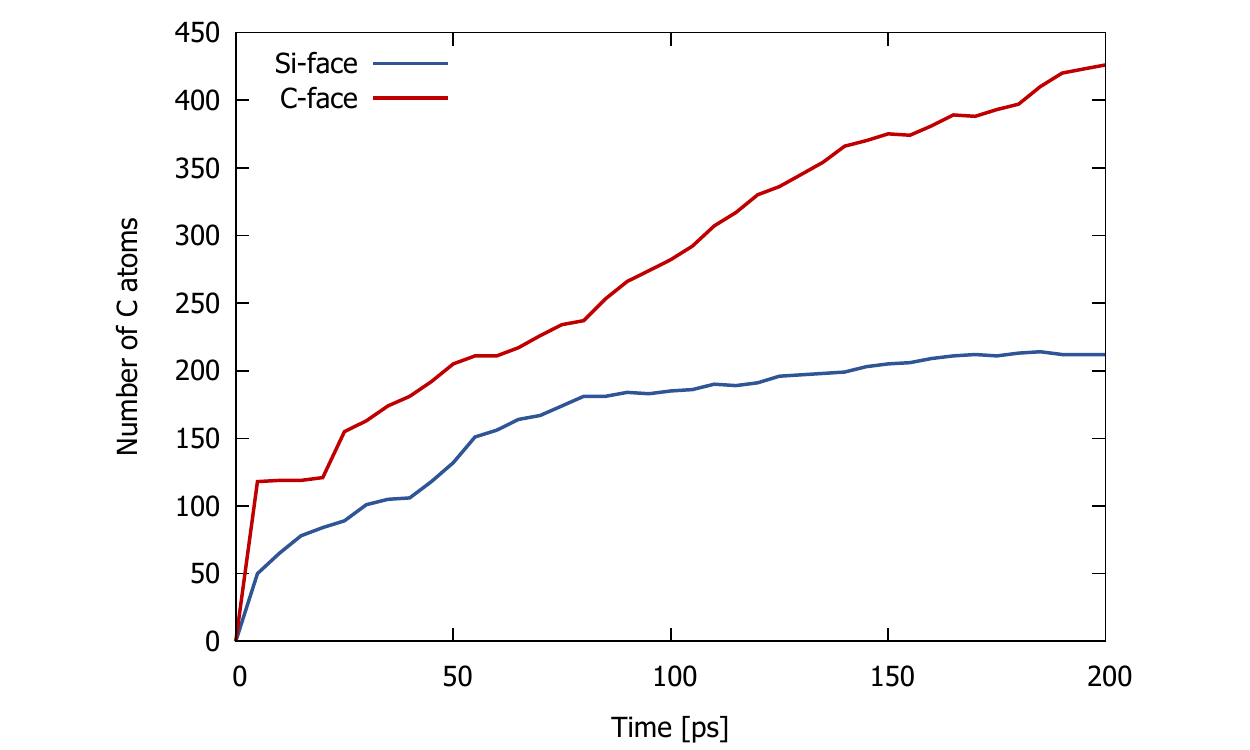}
	\caption{Comparison of the numbers of excess C atoms.}
	\label{fig:num_c_cluster}
\end{figure}

\begin{figure}[tbp]
	\centering
	\begin{minipage}[t]{.40\linewidth}
		\centering
		\includegraphics[width=1.\linewidth,trim=00 00 00 00]{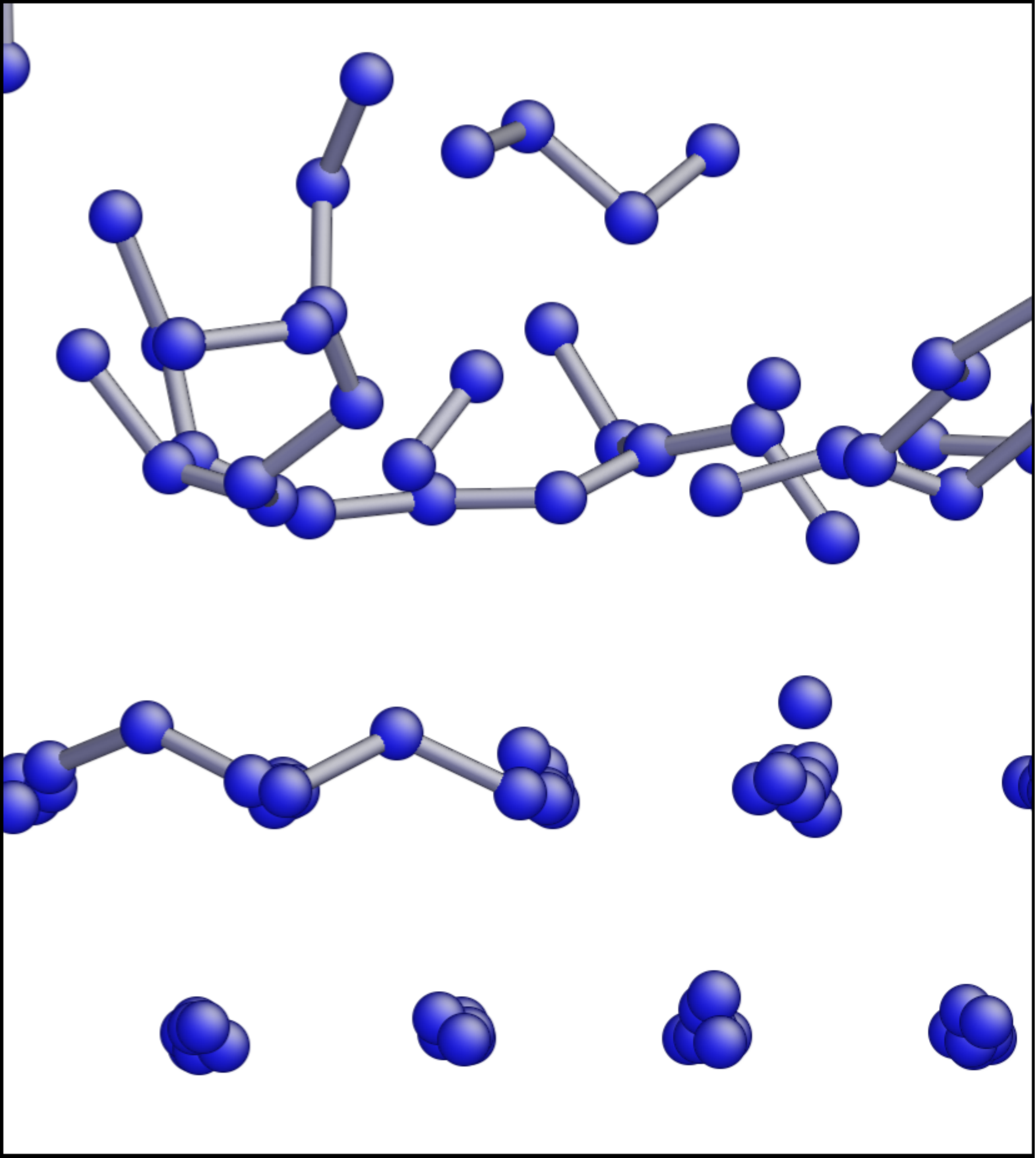}
		\subcaption{Si-face after 140 ps}
	\end{minipage}
	\begin{minipage}[t]{.40\linewidth}
		\centering
		\includegraphics[width=1.\linewidth,trim=00 00 00 00]{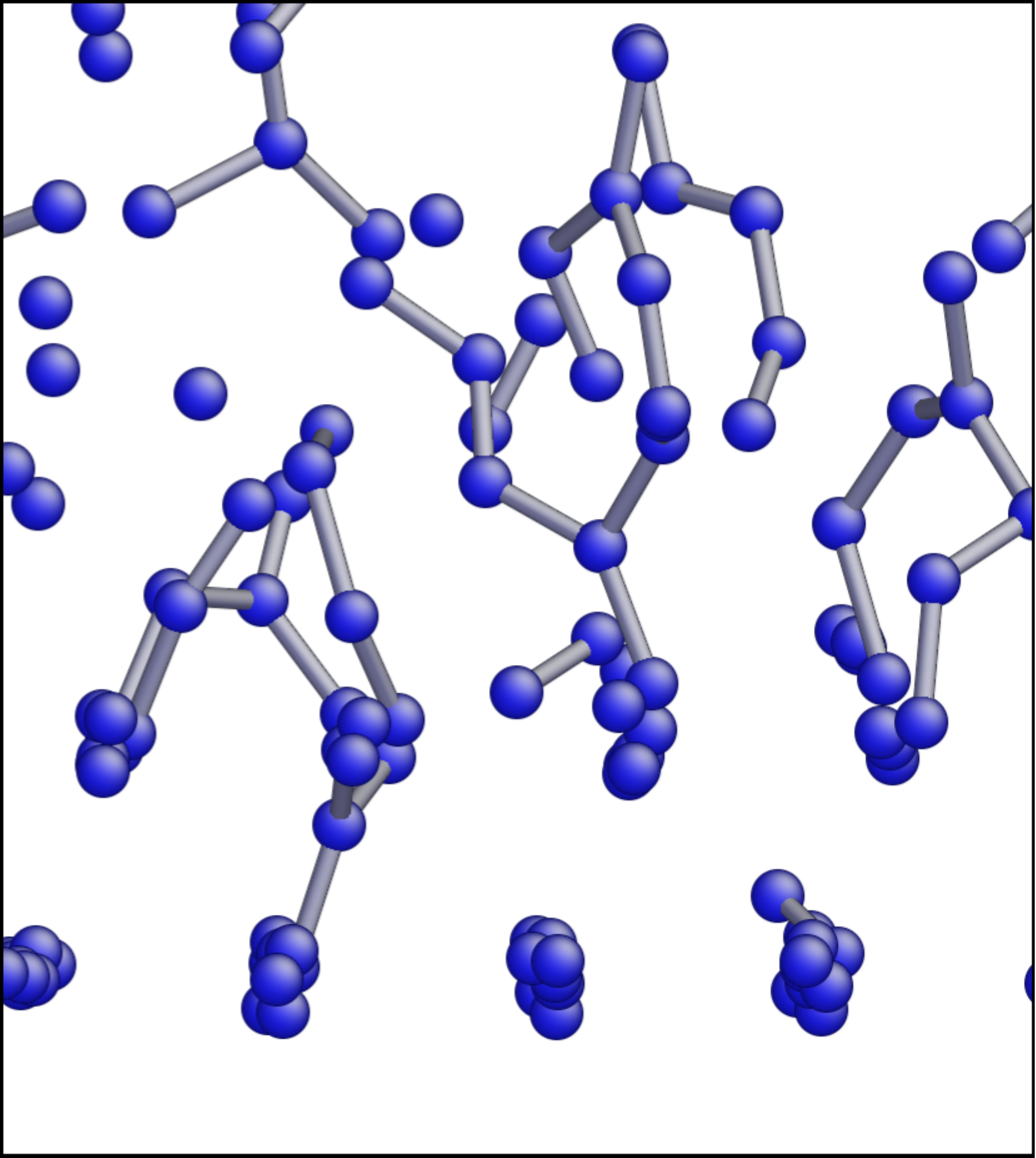}
		\subcaption{C-face after 70 ps}
	\end{minipage}
	\caption{Comparison of the C-C bonds at the SiC/SiO${}_2$ interface. These snapshots correspond to the dashed box shown in \figref{fig:sic_md_oxidation_snapshots_c}.}
	\label{fig:sic_md_oxidation_snapshots_c_enlarged}
\end{figure}

The difference in the oxidation process also affects the creation of excess C atoms at the interface.
Figure \ref{fig:num_c_cluster} shows the time histories of the number of C atoms which have C-C bonds at 1600 K.
Since the oxidation in the case of the C-face proceeds quickly as described above, many C atoms are left in the SiO${}_2$ region.
In addition, the geometric difference in the atomic interface structure is likely to affect the formation of the C-C bonds at the interface.
In the case of the C-face, the pulling up of Si atoms breaks the Si-C back bond and the nearby excess C atoms sometimes enter the broken back-bond site.
This process creates vertical C-C bonds at the interface.
This can be seen in  \figref{fig:sic_md_oxidation_snapshots_c}.
The detailed snapshots of C-C bonds at the interface are shown in \figref{fig:sic_md_oxidation_snapshots_c_enlarged}.

\section{\label{sec:Conclusion}Conclusion}
A new interatomic potential was developed to reproduce the kinetics of the thermal oxidation of SiC.
Many snapshots, including oxidation simulations using DFT calculations, were used in the fitting as the training data.
The newly developed potential was then used to perform large-scale SiC oxidation simulations at various temperatures.
The results showed that the activation energy of the Si-face tended to be much larger than that of the C-face, which was consistent with the experimental results.
It was also found that there was a large difference in the intermediate oxide states of Si atoms at the SiC/SiO${}_2$ interface.
In the case of the Si-face, a flat and aligned interface structure that mainly included Si${}^{1+}$ was created.
The activation energy of the oxidation process of those Si${}^{1+}$ was estimated to be higher than that of the C-face.
Based on these results, we propose that the stability of the flat interface structure is the origin of the high activation energy of the Si-face oxidation.
In contrast, in the case of the C-face, it is found that the Si atoms at the interface are easily pulled up by the O atoms and break the Si-C back bonds.
This process forms the disordered interface and decreases the activation energy of the oxidation.
Since the oxidation in the case of the C-face proceeds quickly, it is also proposed that many excess C atoms are created in the case of the C-face.

\begin{acknowledgments}
A portion of this research was partly supported by MEXT within the priority issue 6 of the FLAGSHIP2020 and JSPS KAKENHI Grant No. 16H03830.

The DFT-MD calculations were carried out on the K computer provided by RIKEN, AICS through the HPCI System Research project (Project ID: hp150266 and hp160226).

This work is supported by a Grant-in-Aid for JSPS Research Fellow.
\end{acknowledgments}

\bibliography{main.bbl}

\begin{thebibliography}{29}
\bibitem [{lam()}]{lammpssource}%
  \BibitemOpen
  \href@noop {} {}\bibinfo {note} {The implementation and parameters are
  available at [URL will be provided after this manuscript is
  accepted].}\BibitemShut {Stop}%
\end{thebibliography}

\end{document}